\documentclass[runningheads]{llncs}
\usepackage{graphicx}
\usepackage{booktabs}
\usepackage{rotating}
\usepackage{url}
\newcommand{\subhead}[1]{\par\vspace {1pt}\noindent{\textbf{#1.}}} 
\newcommand{\formatSpecial}[1]{\texttt{#1}} 
\newcommand*\rot{\rotatebox{90}}
\graphicspath{ {./images/} }
\usepackage{fp} 
\newcommand\printpercent[2]{\FPeval\result{round(#1*100/#2,1)}\result\%} 

\newcommand{\crawlPeriod}{Sept.\ 2018 to Apr.\ 2019} 
\newcommand{\crawlDatasetSize}{18.5GB} 
\newcommand{\crawlDataset}{679} 
\newcommand{\crawlAcceptedDataset}{576} 
\newcommand{\crawlIgnoredDataset}{103} 
\newcommand{\crawlHotspots}{80} 
\newcommand{\crawlUniqueLocations}{67} 
\newcommand{\crawlMultipleLocations}{12} 

\newcommand{\adinjectionlDatasetSize}{8.7GB} 
\newcommand{\adinjectionDataset}{368} 
\newcommand{\adinjectionAcceptedDataset}{368} 
\newcommand{\adinjectionHotspots}{98} 
\newcommand{\adinjectionUniqueHotspots}{87} 
\newcommand{\adinjectionMultipleLocations}{11} 

\newcommand{\hotspotNoneRequirePII}{40} 
\newcommand{\hotspotRequirePII}{27} 
\newcommand{\hotspotMandatoryPII}{19} 
\newcommand{\hotspotsThirdPartyCaptivePortal}{40} 
\newcommand{\readMacaddress}{46} 
\newcommand{\leakMacaddress}{40} 
\newcommand{\redirectToWebsite}{46} 
\newcommand{\HotspotcreateCookiesWithoutConsent}{26} 

\newcommand{\CPAvgtrackingDomains}{7.4} 
\newcommand{\CPMaxgtrackingDomains}{34} 
\newcommand{\CPknowntrackingDomains}{10} 
\newcommand{\LPAvgtrackingDomains}{30.6} 
\newcommand{\CPAvgCookie}{4.2} 
\newcommand{\CPMaxCookie}{34} 
\newcommand{\CPHostpotsCreateCookie}{40} 

\begin{document}
	\title{On Privacy Risks of Public WiFi Captive Portals} 
	\author{Suzan Ali \and
	    Tousif Osman \and
		Mohammad Mannan \and
		Amr Youssef}
	\authorrunning{S.\ Ali et al.}
	\institute{Concordia University,
		Montreal, Canada\\
		\email{\{a\_suzan,t\_osma,mmannan,youssef\}@ciise.concordia.ca}}
	\maketitle 

	\begin{abstract}
		Open access WiFi hotspots are widely deployed in many public places, including restaurants, parks, coffee shops, shopping malls, trains, airports, hotels, and libraries. While these hotspots provide an attractive option to stay connected, they may also track user activities and share user/device information with third-parties, through the use of trackers in their captive portal and landing websites. In this paper, we present a comprehensive privacy analysis of \crawlUniqueLocations{} unique public WiFi hotspots located in Montreal, Canada, and shed some light on the web tracking and data collection behaviors of these hotspots. Our study reveals the collection of a significant amount of privacy-sensitive personal data through the use of social login (e.g., Facebook and Google) and registration forms, and many instances of tracking activities, sometimes even before the user accepts the hotspot's privacy and terms of service policies. Most hotspots use persistent third-party tracking cookies within their captive portal site; these cookies can be used to follow the user's browsing behavior long after the user leaves the hotspots, e.g., up to 20 years. Additionally, several hotspots explicitly share (sometimes via HTTP) the collected personal and unique device information with many third-party tracking domains. 
	\end{abstract}
	
	\section{Introduction}
	Public WiFi hotspots are growing in popularity across the globe. Most users frequently connect to hotspots due to their free-of-cost service, (as opposed to mobile data connections) and ubiquity. According to a Symantec study~\cite{RefWorks:5} conducted among 15,532 users across 15 global markets, 46\% of participants do not wait more than a few minutes before connecting to a WiFi network after arriving at an airport, restaurant, shopping mall, hotel or similar locations. Furthermore, 60\% of the participants are unaware of any risks associated with using an untrusted network, and feel their personal information is safe.
	
	A hotspot may have a captive portal, which is usually used to communicate the hotspot's privacy and terms-of-service (TOS) policies, and collect personal identification information such as name and email for future communications, and authentication if needed (e.g., by asking the user to login to their social media sites). Upon acceptance of the hotspot's policy, the user is connected to the internet and her web browser is often automatically directed to load a landing page (usually the service provider's webpage).
	
	Several past studies (e.g.,~\cite{RefWorks:2,RefWorks:4}) focus on privacy leakage from browsing the internet or using mobile apps in an open hotspot, due to the lack of encryption, e.g., no WPA/WPA2 support at the hotspot, and the use of HTTP, as opposed to HTTPS for connections between the user device and the web service. However, in recent years, HTTPS adoption across web servers has increased dramatically, mitigating privacy exposure through plain network traffic. For example, according to the Google Transparency Report~\cite{https-stats}, as of Apr.\ 6, 2019, 82\% of web pages are served via HTTPS for Chrome users on Windows. On the other hand, in the recent years, there have also been several comprehensive studies on web tracking on regular web services and mobile apps with an emphasis on most popular domains/services (see e.g.,~\cite{RefWorks:7,cross-device,trackers-toit}).

	In contrast to past hotspot and web privacy measurement studies, we analyze tracking behaviors and privacy leakage in WiFi captive portals and landing pages. We design a data collection framework (\texttt{CPInspector}) for both Windows and Android, and capture raw traffic traces from several public hotspots (in Montreal, Canada) that require users to go through a captive portal before allowing internet access. Challenges here include: manual collection of captive portal data by physically visiting each hotspot; making our test environment separate from the regular user environment so that we do not affect the user's browsing profiles; ensuring that our tests remain unaffected by the user's past browsing behaviors (e.g., saved tracking cookies); and creating and monitoring several test accounts in popular social media or email services as some hotspots mandate such authentication. 
	CPInspector does not include any real user information in the collected dataset, or leak such information to the hotspots (e.g., by using fake MAC addresses).

	From each hotspot, we collect traffic using both Chrome and Firefox on Windows. In addition to the default browsing mode, we also use private browsing, and deploy two ad-blockers to check if such privacy-friendly environments help against captive portal trackers---leading to a total of eight datasets for each hotspot. We also use social logins (Facebook, LinkedIn, Google, Instagram, Twitter) if required by the captive portal, or provided as an option; we again use both browsers for social login tests (two to six additional datasets as we have observed at most three social login options per hotspot). Some hotspots also require the user to complete a registration form that collects the user's PII---in such cases, we collect two more datasets (from both browsers). Finally, some hotspots collect additional personal information as part of an optional survey. When reporting statistics on tracking domains and cookies, we accumulate the \emph{distinct} trackers as observed in all the datasets collected for a given hotspot.
	
	On Android, we collect traffic only from the custom captive portal app (as opposed to Chrome/Firefox on Windows) as the cookie store of this app is separate from browsers. Consequently, tracking cookies from the Android captive portal app cannot be used by websites loaded in a browser. Recent Android OSes also use dynamic MAC addresses, limiting MAC address based tracking. However, we found that cookies in the captive portal app may remain valid for up to 20 years, allowing effective tracking by hotspot providers.

	We also design our framework to detect ad/content injection by hotspots; however, we observed no content modification attempts by the hotspots.
	Furthermore, we manually evaluate various privacy aspects of some hotspots, as documented in their privacy/terms-of-service policies, and then compare the stated policies against what happens in practice. {\bf Note:} \emph{by default all our statistics refer to the measurements on Windows; we explicitly mention when results are for Android (mostly in Sec.\ref{sec:android})}.

	\subhead{Contributions and summary of findings}
	\vspace{-10pt}
	\begin{enumerate}
		\item We collected a total of \crawlDataset{} datasets from the captive portal and landing page of \crawlHotspots{} hotspot locations 
		between \crawlPeriod{}. 
		\crawlIgnoredDataset{} datasets were discarded due to some errors (e.g., network failure). We analyzed over \crawlDatasetSize{} of collected traffic for privacy exposure and tracking, and report the results from \crawlUniqueLocations{} unique hotspots (\crawlAcceptedDataset{} datasets), making this
		the largest such study to characterize hotspots in terms of their privacy risks. 
		
		\item Our hotspots include cafes and restaurants, shopping malls, retail businesses, banks, and transportation companies (bus, train and airport), some of which are local to Montreal, but many are national and international brands. \hotspotsThirdPartyCaptivePortal{} hotspots (\printpercent{\hotspotsThirdPartyCaptivePortal}{\crawlUniqueLocations}) use third-party captive portals that appear to have many other business customers across Canada and elsewhere. Thus our results might be applicable to a larger geographical scope. 
		
		\item \hotspotRequirePII{} hotspots (\printpercent{\hotspotRequirePII}{\crawlUniqueLocations}) use social login or a registration page to collect personal information (\hotspotMandatoryPII{} hotspots make this process mandatory for internet access). Social login providers may share several privacy-sensitive PII items---e.g., we found that LinkedIn shares the user's full name, email address, profile picture, full employment history, and the current location.

		\item Except three, all hotspots employ varying levels of user tracking technologies on their captive portals and landing pages. On average, we found \CPAvgtrackingDomains{} third-party tracking domains per captive portal (max: \CPMaxgtrackingDomains{} domains). 
		\CPHostpotsCreateCookie{} hotspots (\printpercent{\CPHostpotsCreateCookie}{\crawlUniqueLocations}) create persistent third-party tracking HTTP cookies (validity up to 20 years); \CPAvgCookie{} cookies on average on each captive portal (max: \CPMaxCookie{} cookies). Surprisingly, \HotspotcreateCookiesWithoutConsent{} hotspots (\printpercent{\HotspotcreateCookiesWithoutConsent}{\crawlUniqueLocations}) create persistent cookies even \emph{before} getting  user consent on their privacy/TOS document.

		\item Several hotspots explicitly share (sometimes even without HTTPS) personal and unique device information with many third-party domains. \leakMacaddress{} hotspots (\printpercent{\leakMacaddress}{\crawlUniqueLocations}) expose the user's device MAC address; five hotspots leak PII via HTTP, including the user's full name, email address, phone number, address, postal code, date of birth, and age (despite some of them claiming to use TLS for communicating such information). Two hotspots appear to perform cross-device tracking via Adobe Marketing Cloud Co-op~\cite{RefWorks:74}. 
		
		\item Two hotspots (\printpercent{2}{\crawlUniqueLocations}) state in their privacy policies that they explicitly link the user's MAC address to the collected PII, allowing long-term user tracking, especially for desktop OSes with fixed MAC. 
		
		\item From our Android experiments, we reveal that 9 out of 22 hotspots can effectively track Android devices even though Android uses a separate captive portal app and randomizes MAC address as visible to the hotspot.

	\end{enumerate}
	
	\vspace{-5pt}
	\section{Background and Related Work}
	\vspace{-5pt}
	In this section, we first provide an overview of public hotspots, and then briefly review related previous studies on hotspots, web tracking, and ad injection.
	
	Hotspot access is usually deployed in three forms: captive portal, direct/open-access (no captive portals), or password-protected networks. In captive portal networks, users first go through a captive portal session before getting internet access. The captive portal web-page usually displays the privacy policy and/or the terms-of-service (TOS) document, along with some advertisements, and sometimes an option to select their preferred language (for viewing the portal content), and a social login or registration form. After accepting the policy/TOS documents, the user's browser is often directed to a \emph{landing} page, as chosen by the hotspot owner. The captive portal is used to make sure that guests are aware of the hotspot privacy policy, collect personal identification information such as name and email for future communications, and authenticate guests if needed. 
	
	Several prior studies have demonstrated the possibility of eavesdropping WiFi traffic to identify personal sensitive information in public hotspots. For example, Cheng et al.~\cite{RefWorks:2} collected WiFi traffic from 20 airports in four countries, and found that two thirds of the travelers leak private information while using airport hotspots for web browsing and smartphone app usage. Sombatruang et al.~\cite{RefWorks:4} conducted a similar study in Japan by setting up 11 experimental open public WiFi networks. The 150 hour experiment confirmed the exposure of private information, including photos, email addresses, confidential documents, and users' credentials---transmitted via HTTP. In contrast, we analyze web tracking and privacy leakage within WiFi captive portals and landing pages. Klasnja et al.~\cite{klasnja2009wi} studied privacy and security awareness of WiFi users by monitoring web traffic of 11 users.  
	The study shows the users' limited understanding of risks associated with WiFi usage, and a false sense of safety.
	
	Web tracking, a widespread phenomenon on the internet, is used for varying purposes, including: targeted advertisements, identity checking, website analytics, and personalization. Web tracking techniques can generally be categorized as stateful and stateless. Eckersley~\cite{eckersley2010unique} showed that 83.6\% of the Panopticlick website~\cite{RefWorks:107} visitors could be uniquely identified from a fingerprint composed of only 8 attributes.
	Laperdrix et al.~\cite{laperdrix2016beauty} showed that \url{AmIUnique.org} can uniquely identify 89.4\% of fingerprints composed of 17 attributes, including the HTML5 canvas element and the WebGL API. In a more recent large-scale study, G{\'o}mez-Boix et al.~\cite{gomez2018hiding} collected over 2 million real-world device fingerprints (composed of 17 attributes) from a top French website; they found that only 33.6\% device fingerprints are unique, raising questions on the effectiveness of fingerprinting in the wild. Note that developing advanced fingerprinting techniques to detect the so-called \emph{golden image} (the same software and hardware as often deployed in large enterprises), is an active research area---see e.g.,~\cite{dns-fp,hwclock-fp}.   
	Several automated frameworks have also been designed for large-scale measurement of web tracking in the wild; see e.g., FPDetective~\cite{RefWorks:34} and OpenWPM~\cite{RefWorks:7}. 
	In this work, we measure tracking techniques in captive portals and landing pages, and use OpenWPM to verify the prevalence of the found trackers on popular websites.

	Previous work has also looked into ad injection in web content, see e.g.,~\cite{RefWorks:70,RefWorks:73}.
	We use similar methods for detecting potential similar content injection in hotspots since such incidents have been reported in the past (e.g.,~\cite{RefWorks:65,RefWorks:66,RefWorks:67}).
	
	\vspace{-5pt}
	\section {CPInspector on Windows: Design and Data Collection}
	\label{CPInspector_Framework} 
\begin{figure}[ht]
	\centering
	\vspace{-18pt}
\includegraphics[scale=0.75, angle =90, trim={10cm 1cm 10cm 5.3cm}, clip ]{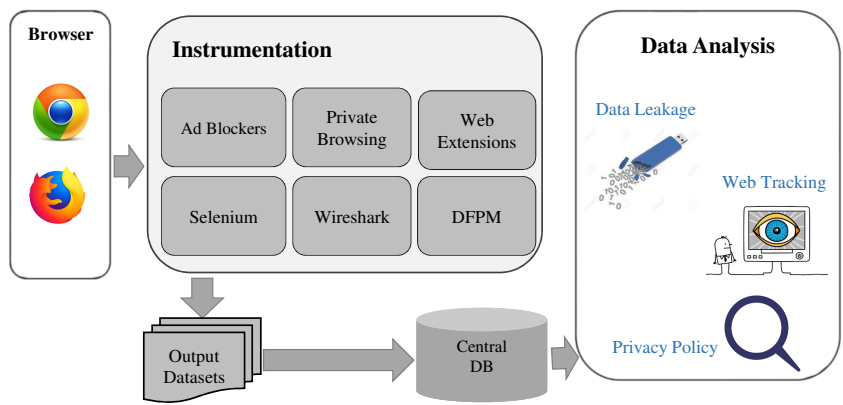} 
	\vspace{-15pt}
		\caption{CPInspector components}
		\label{fig:architecturefig}
		\vspace{-15pt}
	\end{figure}

	\vspace{-5pt}
	In this section, we describe CPInspector, the platform we develop for measuring captive portal web-tracking and privacy leakages; see Fig.~\ref{fig:architecturefig} for the Windows variant. As Android uses a special app for captive portal, we modify CPInspector accordingly; see Sec.~\ref{sec:android}. 
	
	The main components of CPInspector include: a browser automation framework, a data migration tool and an analysis module. Our browser automation platform, which is driven by Selenium~\cite{RefWorks:3}, is used to visit the hotspot captive portal and perform a wide range of measurements. It collects web traffic, HTTP cookies, WebStorage content, fingerprints, browsing profiles, page source code, and screen shots of rendered pages (used to verify the data collection process). It also saves a copy of the privacy policy, if found. The datasets collected from our evaluated hotspots are parsed and committed to a central SQLite database. CPInspector works on Windows 7/8/10 with Firefox Quantum v60.1.0ESR and Google Chrome v69.0.3497.100 browsers. It is developed using Selenium, Wireshark 2.6.2+, Node.js 8.1.1.4, WebExtensions, and Python 3.7+. 
	
	\subhead{Capturing traffic}
	We use Wireshark~\cite{RefWorks:26} to capture all traffic between the instrumented browser and the hotspot access point. We filter out traffic generated by normal activities such as anti-virus scanning and Windows updates. Moreover, since some captive portals adopt TLS for communication, we rely on the SSLKEYLOGFILE~\cite{RefWorks:23} to decrypt the TLS traffic; we then use Tshark~\cite{RefWorks:24} to extract and save the HTTP requests/responses to our database.
	
	\subhead{Identifying third-parties}
	We identify third-party domains using the hotspot website's owner. All evaluated hotspots have an official website except Hvmans Cafe, where all domains are classified as third-parties. We primarily use the Python WHOIS library~\cite{RefWorks:41} to find domain registration information. In cases where the domain information is protected by the WHOIS privacy policy, we visit the domain to detect any redirect to a parent site; we then lookup the parent site's registration information. If this fails, we manually review the domain's \texttt{Organization} in its TLS certificate, if available. Otherwise, we try to identify the domain owner based on its WHOIS registration email; e.g., \url{addthis.com} is owned by Oracle as apparent from its WHOIS email \url{domain-contact_ww_grp@oracle.com}. We also use Crunchbase~\cite{RefWorks:42} and Hoovers~\cite{RefWorks:43} to determine if the organizations are subsidiaries or acquisitions of larger companies; e.g., \url{instagram.com} is owned by Instagram, which in turn is owned by Facebook.

	\subhead{Identifying third-party trackers}
	We use EasyList~\cite{RefWorks:30} with EasyPrivacy, and Fanboy's List to identify known third-party trackers. EasyList identifies known advertising-related trackers, EasyPrivacy detects known non-advertising-related trackers, and Fanboy's list classifies known social media content related trackers. These lists rely on blacklisted script names, URLs, or domains, which may fail to detect new trackers or variations of known trackers. 
	For this reason, we classify third-party trackers as follows: 
	(a) A \emph{known tracker} is a third-party that has already been identified in the above blacklists.
	(b) A \emph{possible tracker} is any third-party that can potentially track the user's browsing activities but not included in a blacklist.
	We observed variations of well-known trackers such as Google Analytics, were missed by the blacklists (see Table~\ref{fig:Possible_Trackers} in the appendix).

	\subhead{Email and social login accounts}
	We registered 27 accounts in total to be used in our experiments (if required by the captive portal), including: 4 Gmail, 1 Yahoo, 3 Microsoft, 8 Facebook, 4 Instagram, 6 LinkedIn, and 1 Twitter. 
	
	\subhead{Ad injection detection}
	Our framework also includes a module to detect modifications to user traffic, e.g., for ad injections. We visit two decoy websites (i.e., honeysites in our control), via a home network and a public hotspot, and then compare the differences in the retrieved content (i.e., DOM trees~\cite{RefWorks:69}). The use of honeysites allows us to avoid any false positive issues due to the website's dynamic content (e.g., news updates, dynamic ads). However, we also include a real website in our experiments (\url{BBC.com}). The first honysite is a static web page while the second is comprised of dynamic content that incorporates JavaScript elements, iframe tags, and four fake ads. The fake ads were created based on source code snippets from Google Adsense~\cite{RefWorks:86}, Google TagManager~\cite{RefWorks:85}, Taboola~\cite{RefWorks:87}, and BuySellAds~\cite{RefWorks:88}. We host the honeysites through Amazon AWS and carefully mimic a realistic website.

	\subhead{Data collection}
	We collected a total of \crawlDataset{} datasets from the captive portal and landing page of \crawlHotspots{} hotspots (\crawlMultipleLocations{} hotspots are measured at multiple physical locations) between \crawlPeriod{}. We stopped collecting datasets from different locations of the same chain-business as the collected datasets were largely the same. We discarded \crawlIgnoredDataset{} datasets due to some errors (e.g., network failures, incomplete data collection scenarios). We analyzed over \crawlDatasetSize{} of collected traffic for privacy exposure and tracking measurements, and report the results from \crawlUniqueLocations{} unique hotspots (\crawlAcceptedDataset{} datasets). We discuss the results in Sec.~\ref{sec:cp-results}.

	For the ad injection experiments, we collected a total of \adinjectionDataset{} datasets from crawling the two honey websites and the \url{BBC.com} website at \adinjectionHotspots{} hotspots; \adinjectionMultipleLocations{} hotspots are measured at multiple physical locations. 
	We analyzed over \adinjectionlDatasetSize{} of collected traffic for ad injection, and report the results from \adinjectionUniqueHotspots{} unique hotspots (\adinjectionAcceptedDataset{} datasets). We did not observe any content modification attempts.

 	\section{Analysis and Results for Windows}\label{sec:cp-results}
	In this section, we present the results of our analysis on collected personal information, privacy leaks, web trackers, HTTP cookies, fingerprinting, and the effectiveness of two anti-tracking extensions and private browsing mode.
	
    \vspace{-8pt}	
	\subsection{Personal Information Collection, Sharing, and Leaking}
	\vspace{-4pt}
	\subhead{PII collection}
	Most hotspots (\hotspotNoneRequirePII; \printpercent{\hotspotNoneRequirePII}{\crawlUniqueLocations}) allow internet access without seeking any explicit personal data. The remaining \hotspotRequirePII{} hotspots use social login (Facebook, LinkedIn, Google, Instagram), or a registration page to collect significant amount of personal information; \hotspotMandatoryPII{} of these hotspots mandate social login or user registration. An optional survey is also used by one hotspot. 
	See Table~\ref{table:personalinfotable}.  
	
	\begin{table}[!t]
		\caption{Personal information collected via social login, registration, or optional surveys. The ``Powered By'' column refers to third-parties that provide hotspot services (when used/identified). \formatSpecial{F} refers to Facebook, \formatSpecial{L}: LinkedIn, \formatSpecial{I}: Instagram, \formatSpecial{G}: Google, \formatSpecial{T}: Twitter, \formatSpecial{R}: registration form, and \formatSpecial{S}: survey; *: personal information is mandatory to access the service.} 
		\label{table:personalinfotable}
		\vspace{-8pt}
		\resizebox{\textwidth}{!}{%
			
			\begin{tabular}{@{}llccccccccccccccccccc@{}}
				\textbf{\textbf{Hotspot}} & \textbf{Powered By} & \textbf{\textbf{\rot{Name}}} & \textbf{\textbf{\rot{Email}}} & \textbf{\textbf{\rot{Gender}}} & \textbf{\textbf{\rot{Birthday}}} & \textbf{\textbf{\rot{Phone Number}}} & \textbf{\textbf{\rot{Current City}}} & \textbf{\textbf{\rot{Profile Picture}}} & \textbf{\textbf{\rot{Home Town}}} & \textbf{\textbf{\rot{Country}}} & \textbf{\textbf{\rot{Facebook Likes}}} & \textbf{\textbf{\rot{Facebook Friends}}} & \textbf{\textbf{\rot{LinkedIn Headline}}} & \textbf{\textbf{\rot{Current Employment}}} & \textbf{\textbf{\rot{Postal Code}}} & \textbf{\textbf{\rot{\# of Children}}} & \textbf{\textbf{\rot{Basic Profile}}} & \textbf{\textbf{\rot{Instagram Media}}} & \textbf{\textbf{\rot{Tweets}}} & \textbf{\textbf{\rot{People You Follow}}}\\ \midrule
				Bombay Mahal Thali* & Sy5 & \formatSpecial{FR} & \formatSpecial{FR} & & \formatSpecial{F} & & \formatSpecial{F} & & & & & & & & & & & & & \\
				Carrefour Laval* & Aislelabs & \formatSpecial{FR} & \formatSpecial{FR} & \formatSpecial{FR} & \formatSpecial{F} & & \formatSpecial{F} & \formatSpecial{F} & \formatSpecial{F} & & \formatSpecial{F} & & & & & & & & &\\
				Fairview Pointe-Claire* & Aislelabs & \formatSpecial{FR} & \formatSpecial{FRT} & \formatSpecial{FR} & \formatSpecial{F} & & \formatSpecial{F} & \formatSpecial{F} & \formatSpecial{F} & & \formatSpecial{F} & & & & & & & & \formatSpecial{T}& \formatSpecial{T} \\
				Carrefour Angrignon & Eye-In & \formatSpecial{FGL} & \formatSpecial{FGL} & & & & & \formatSpecial{FGL} & & & & & \formatSpecial{L} & \formatSpecial{L} & & & \formatSpecial{L} & & &\\
				Centre Eaton & Eye-In & \formatSpecial{F} & \formatSpecial{F} & & & & & \formatSpecial{F} & & & & & & & & & & & &\\
				Centre Rockland & Eye-In & \formatSpecial{FL} & \formatSpecial{FL} & & & & & \formatSpecial{FL} & & & & & \formatSpecial{L} & \formatSpecial{L} & & & \formatSpecial{L} & & &\\
				Desjardins 360* & JoGoGo & \formatSpecial{F} & \formatSpecial{F} & \formatSpecial{F} & \formatSpecial{F} & \formatSpecial{R} & & \formatSpecial{F} & & & & \formatSpecial{F} & & & & & & & &\\
				Domino's Pizza & & & \formatSpecial{R} & & & & & & & & & & & & & & & & &\\
				Dynamite* & & & \formatSpecial{R} & & & & & & & & & & & & & & & & &\\
				GAP & & & \formatSpecial{R} & & & & & & & & & & & & & & & & &\\
				Garage* & & & \formatSpecial{R} & & & & & & & & & & & & & & & & &\\
				Grevin Montreal & Eye-In & \formatSpecial{FL} & \formatSpecial{FL} & & & & \formatSpecial{F} & \formatSpecial{FL} & & & & & \formatSpecial{L} & \formatSpecial{L} & \formatSpecial{S} & \formatSpecial{S} & \formatSpecial{L} & & &\\
				Harvey's* & Colony Networks & \formatSpecial{F} & \formatSpecial{FR} & & & & & \formatSpecial{F} & & & & & & & & & & & &\\
				Hvmans Cafe* & Purple & \formatSpecial{FR} & \formatSpecial{FR} & & \formatSpecial{F} & & \formatSpecial{F} & \formatSpecial{F} & & & \formatSpecial{F} & & & & & & \formatSpecial{I} & \formatSpecial{I} & &\\
				Mail Champlain & Eye-In & \formatSpecial{FL} & \formatSpecial{FL} & & & & & \formatSpecial{FL} & & & & & \formatSpecial{L} & \formatSpecial{L} & & & \formatSpecial{L} & & &\\
				Maison Simmon* & & & \formatSpecial{R} & & & & & & & & & & & & & & & & &\\
				Michael Kors* & Purple & \formatSpecial{R} & \formatSpecial{R} & \formatSpecial{R} & \formatSpecial{R} & \formatSpecial{R} & & & & & & & & & \formatSpecial{R} & & & &\\
        Montreal Science Centre* & Telus & & \formatSpecial{R} & & & & & & & & & & & & & & & & &\\
				Moose BAWR* & Sticky WiFi & & \formatSpecial{R} & & & & & & & & & & & & & & & & &\\
				Nautilus Plus* & & & \formatSpecial{R} & & & & & & & & & & & & & & & & &\\
				Nespresso* & Orange & & & & & \formatSpecial{R} & & & & & & & & & & & & & &\\
				Place Montreal Trust & & & \formatSpecial{R} & & & & & & & \formatSpecial{R} & & & & & \formatSpecial{R} & & & & &\\
				Roots* & Yelp WiFi & \formatSpecial{R} & \formatSpecial{R} & & & & & & & & & & & & & & & &\\
				Telus* & & & \formatSpecial{R} & & & & & & & & & & & & & & & & &\\
				Sushi STE-Catherine* & MyWiFi & & \formatSpecial{R} & & & & & & & & & & & & & & & & &\\
				Vua Sandwiches* & Coolblue & \formatSpecial{FR} & \formatSpecial{FR} & & & \formatSpecial{R} & & \formatSpecial{F} & & & & & & & & & & & &\\
				YUL Airport* & Datavalet & \formatSpecial{FL} & \formatSpecial{FRL} & & & & & \formatSpecial{FL} & & & & & \formatSpecial{L} & \formatSpecial{L} & & & \formatSpecial{L} &  & &\\ \bottomrule
		\end{tabular}}
		\vspace{-15pt}	
	\end{table}
	The Hvmans Cafe hotspot reads the user's profile information and media from Instagram; the profile may include: the user's email address, mobile phone number, user ID, full name, gender, biography, website, and profile picture. 
	Even after login via Instagram, the user must also complete a form to provide her first name, last name and email address. An option is also given to the user to choose a password for the hotspot. However, there is no login screen to use this password in future visits.
	Similarly, users must create an account at Michael Kors, where they can use the account email/password to login in future visits. In addition, Nespresso requires an activation code sent via SMS. Five hotspots support single sign-on via LinkedIn (Carrefour Angrignon, Mail Champlain, Centre Rockland, and Grevin Montreal). 
 LinkedIn shares the user's full name, email address, profile picture, LinkedIn headlines, current employment, and basic profile consisting of a large list of PII items, including full employment history, and the current location~\cite{RefWorks:58}.  
	
	By analyzing the used email and social login accounts, we found no activities related to the hotspots on social accounts while 5 hotspots used emails to send promotional messages (Dynamite, GAP, Garage, Telus, and Place Montreal Trust). Email is also used to activate WiFi access in YUL Airport, Telus, Hvmans Cafe, and Montreal Science Centre.

	\subhead{Sharing with third-parties}
  Most hotspots share PII and browser/device information with third-parties via the referrer header, the request-URL, HTTP cookie or WebStorage. 
  We identified \hotspotsThirdPartyCaptivePortal{} hotspots (\printpercent{\hotspotsThirdPartyCaptivePortal}{\crawlUniqueLocations}) that use third-party captive portals where they share PII, including 18 (\printpercent{18}{\crawlUniqueLocations}) share email address;  15 (\printpercent{15}{\crawlUniqueLocations})  share user's full name; 12 (\printpercent{12}{\crawlUniqueLocations}) share profile picture; 5 (\printpercent{5}{\crawlUniqueLocations}) share birthday, current city, current employment and LinkedIn headline; see Table~\ref{table:personalinfotable}.  We also found some hotspots' captive portals leak device/browser information to third-parties, including \leakMacaddress (\printpercent{\leakMacaddress}{\crawlUniqueLocations}) leak  MAC address and last visited site; 18 (\printpercent{18}{\crawlUniqueLocations}) leak screen resolution; 26 (\printpercent{26}{\crawlUniqueLocations}) leak user agent; 24 (\printpercent{24}{\crawlUniqueLocations}) leak browser Information and language; and 15 (\printpercent{15}{\crawlUniqueLocations}) leak plugins. Moreover, some hotspots leak the MAC address to multiple third-parties, e.g., Pizza Hut to 11 domains, and H\&M Place Montreal Trust and Discount Car Rental to six third-parties each. Top organizations that receive the MAC addresses include: \url{Network-auth.com} from 21 hotspots,
			Alphabet 18,
			\url{Openh264.org} 12,
			Facebook 10, Datavalet 8,
		 and Amazon 6.

	\subhead{PII leaks via HTTP}
	We search for PII items of our used accounts in the collected HTTP Wireshark traffic, and record the leaked information, including the HTTP request URL, and source (captive portal vs.\ landing page). 
	Three hotspots transmit the user's full name via HTTP (Place Montreal Trust, Nautilus Plus and Roots). In Place Montreal Trust, the user's full name is saved in a cookie (valid for five years), and each time the user connects to the captive portal, the cookie is automatically transmitted via HTTP. Moreover, three hotspots leak the user's email address via HTTP (Dynamite, Roots, and Garage). 
	In Nautilus Plus, a user must enter her membership number in the captive portal. For partially entered membership numbers, the captive portal verifies the identity by displaying personal information of five people in a scrambled way (first and last names,  postal codes, ages, dates of birth, and phone numbers), over HTTP. The user then chooses the right combination corresponding to her personal information. We also confirmed that some of this data belongs to real people by authenticating to this hotspot using ten randomly generated partial membership numbers. Then, we used the reverse lookup in \url{canada411.ca} to confirm the correlation between the returned phone numbers, names, and addresses.
	
	\vspace{-5pt}
	\subsection{Presence of Third-Party Tracking Domains and HTTP Cookies}
\begin{figure*}[ht]
		\includegraphics[width=\textwidth, trim={1.8cm 18cm 3.2cm 2.7cm}, clip ]{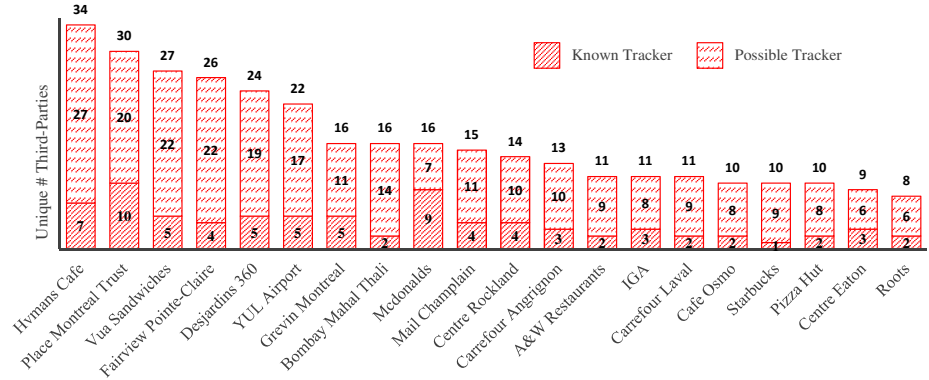}
		\vspace{-25pt}
		\caption{Unique number of third-parties on captive portals (top 20). For example, \mbox{Hvmans} Cafe hosts a total of 34 tracking domains, including 7 known trackers. Note that for all reported tracking/domain statistics, we accumulate the distinct trackers as observed in all the datasets collected for a given hotspot. For list of evaluated hotspots see Table~\ref{hotspotscategory} in the appendix.}
		\label{fig:CP_hotspottracking}
		\vspace{-15pt}
	\end{figure*}

\begin{figure}[ht]
		\includegraphics[width=\textwidth, trim={0.7cm 9.8cm 0.7cm 10.3cm}, clip ]{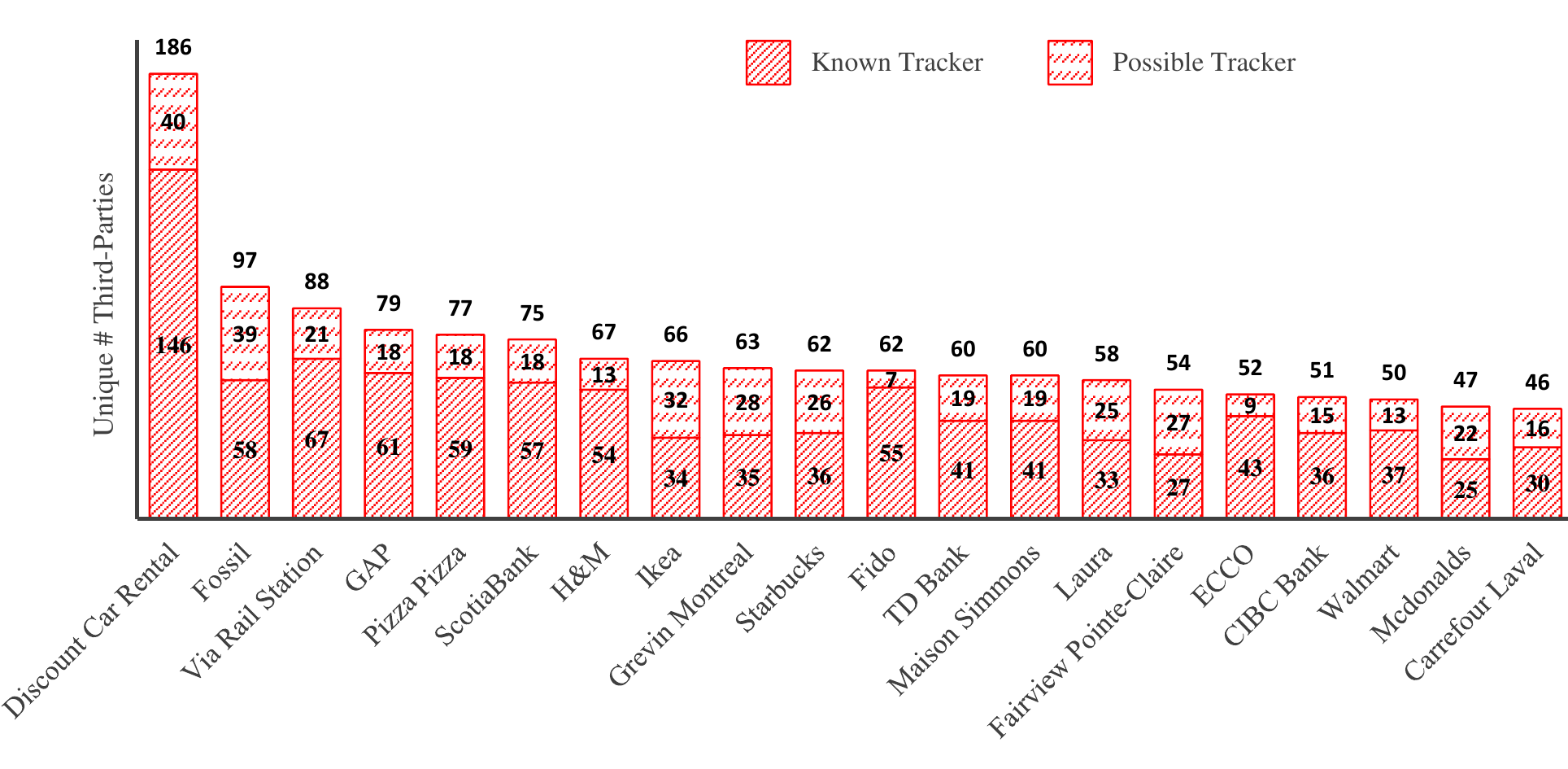}
		\vspace{-30pt}
		\caption{Unique number of third-parties on landing pages (top 20)}
		\label{fig:LP_hotspottracking}
		\vspace{-15pt}
	\end{figure}

	\vspace{-5pt}
	\subhead{Tracking domains}
	We detect third-party tracking domains using: EasyList, EasyPrivacy, and Fanboy's List. On average, each captive portal hosts \CPAvgtrackingDomains{} third-party tracking domains (max: \CPMaxgtrackingDomains{} domains, including \CPknowntrackingDomains{} known trackers); see Fig.~\ref{fig:CP_hotspottracking}. We noticed that the hotspots that use the same third-party captive portal still have a different number of third-parties. For example,  for the Datavalet~\cite{RefWorks:77} hotspots (YUL Airport, McDonald's, Starbucks, Via Rail Station, Tim Hortons, CIBC Bank, Place Vertu), the number of third-parties are 22, 16, 10, 8, 5, 5, and 2 respectively.
	The hotspots (\redirectToWebsite; \printpercent{\redirectToWebsite}{\crawlUniqueLocations}) that redirect users to their corporate websites, host more known third-party tracking domains---on average, \LPAvgtrackingDomains{} domains per landing page; see Fig.~\ref{fig:LP_hotspottracking}. 
	We also analyzed the organizations with the highest known-tracker representations. We group domains by the larger parent company that owns these domains. 
	Alphabet, Facebook, and Datavalet are present on over 10\% of the captive portals. Alphabet and Facebook are also present on over 50\% of the landing pages.
	

	\begin{figure}[ht]
	\vspace{-10pt}
		\includegraphics[width=\textwidth, trim={1.8cm 17.1cm 3.2cm 2.2cm}, clip ]{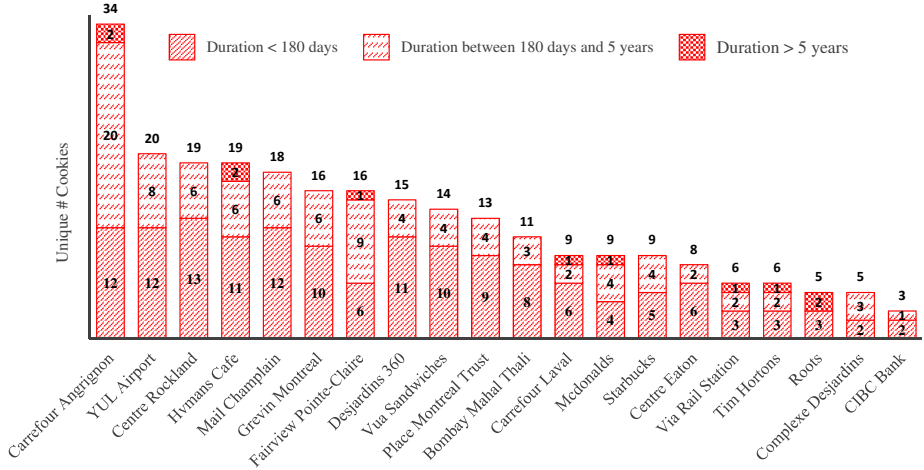}
		\vspace{-20pt}
		\caption{Number of third-party cookies on captive portals (top 20). Note that for all reported cookies/domain statistics, we accumulate the distinct cookies as observed in all the datasets collected for a given hotspot.}
		\label{fig:CP_hotspotthirdpartycookies} 
	\vspace{-20pt}
	\end{figure}
	
	\begin{figure}[ht]
	
		\includegraphics[width=\textwidth, trim={1.7cm 17.5cm 3.2cm 2.2cm}, clip ]{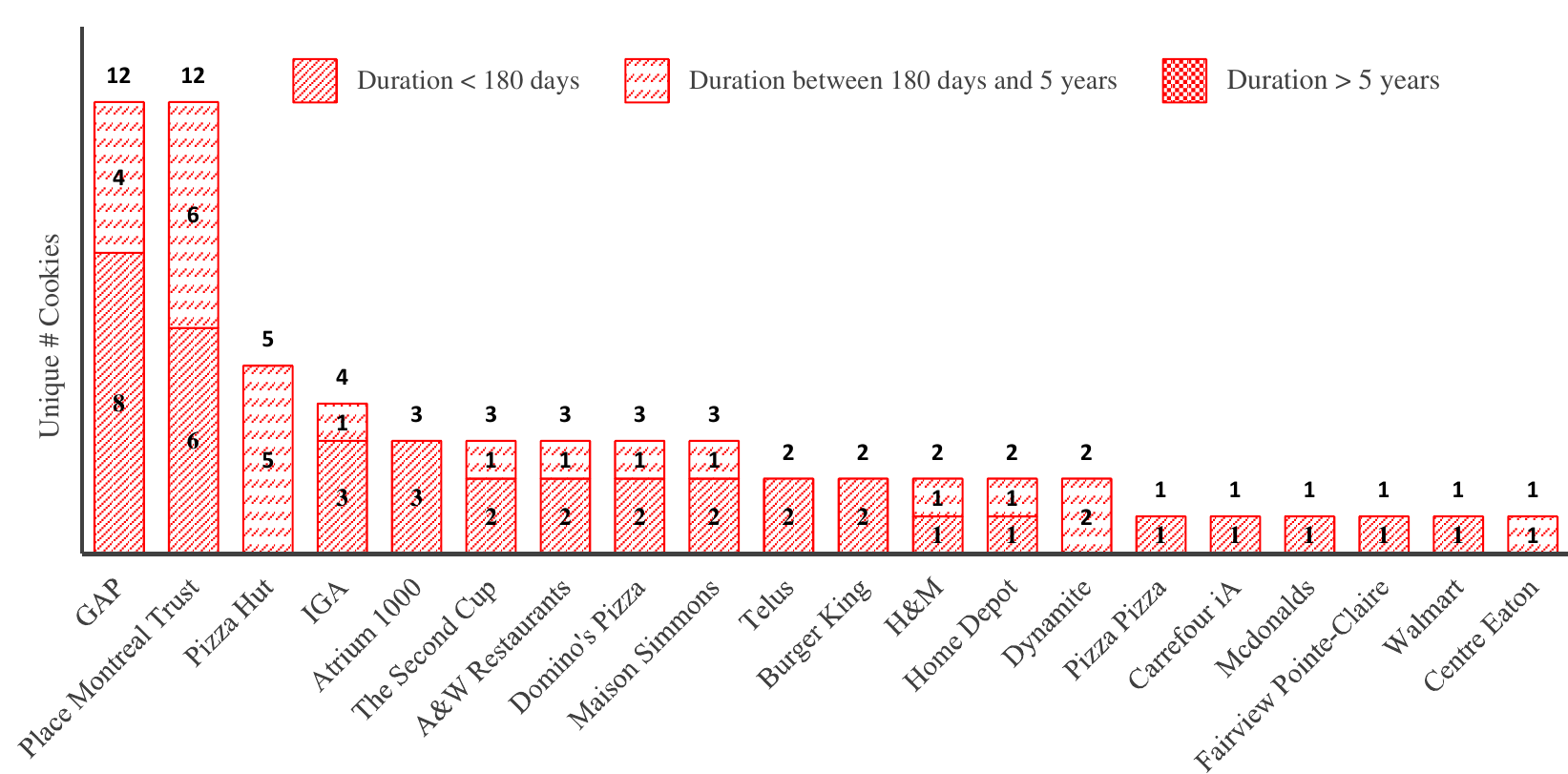}
		\vspace{-20pt}
		\caption{Number of first-party cookies on captive portals (top 20)}
		\label{fig:CP_hotspotfirstpartycookies} 
		\vspace{-20pt}
	\end{figure}

	\subhead{HTTP tracking cookies on captive portals}
	We found \CPHostpotsCreateCookie{} (\printpercent{\CPHostpotsCreateCookie}{\crawlUniqueLocations}) hotspots create third-party cookies valid for various duration---e.g., over 5 years from   10 (\printpercent{10}{\crawlUniqueLocations}) hotspots, six months to five years from 23 (\printpercent{23}{\crawlUniqueLocations}) hotspots, and under six months from 38 (\printpercent{38}{\crawlUniqueLocations}) hotspots; see Fig.~\ref{fig:CP_hotspotthirdpartycookies}. Via Rail Station, Fairview Pointe-Claire, Carrefour Laval, Roots, McDonald's, Tim Hortons, and Harvey's have a third-party cookie from \url{network-auth.com}, valid for 20 years. 
	Moreover, YUL Airport, Via Rail Station, Complexe Desjardins, McDonald's, Starbucks, Tim Hortons, CIBC Bank have a common 1-year valid cookie from Datavalet, except for CIBC (17 days). 
	This cookie uniquely identifies a device based on the MAC address (set to the same value unless the MAC address is spoofed). 
	Some hotspots save the MAC address in HTTP cookies, including CHU Sainte-Justine, Moose BAWR, and Centre Rockland. 

	We also analyze first-party cookies on captive portals; see Fig.~\ref{fig:CP_hotspotfirstpartycookies}.  22 (\printpercent{22}{\crawlUniqueLocations}) hotspots create first-party cookies valid for various durations; 14 (\printpercent{14}{\crawlUniqueLocations}) hotspots include cookies valid for periods ranging from six months to five years, and 17 (\printpercent{17}{\crawlUniqueLocations}) hotspots for less than 6 months. Place Montreal Trust saves the user's full name in a first-party cookie valid for five years; this cookie is transmitted via HTTP.
	Finally, we analyzed hotspots that create persistent cookies before explicit consent from the user, we found \HotspotcreateCookiesWithoutConsent{} (\printpercent{\HotspotcreateCookiesWithoutConsent}{\crawlUniqueLocations}) hotspots  create cookies that are valid for periods varying from  30 minutes to a year, including Domino's Pizza, Fido, GAP, H\&M, McDonald's, Roots, Starbucks, and Tim Hortons.

	\subhead{HTTP tracking cookies on landing pages}
	We found 48 (\printpercent{48}{\crawlUniqueLocations}) hotspots create third-party cookies valid for various durations---e.g., over 5 years from 4 (\printpercent{4}{\crawlUniqueLocations}) hotspots, six months to five years from   47 (\printpercent{47}{\crawlUniqueLocations})  hotspots, and under six months from 42 (\printpercent{42}{\crawlUniqueLocations})  hotspots; see Fig.~\ref{fig:LP_hotspotthirdpartycookies} in the appendix. Prominent examples include the following. Fossil has a 25-year valid cookie from \url{pbbl.com}; CIBC Bank has two 5-year valid cookies from \url{stackadapt.com}, a known tracker. 
	Moreover, H\&M, Starbucks, Laura, Fido, Gap Canada, Harvey's, and Ikea have multiple ten-year valid first-party cookies, but their names suggest a relationship with Optimizely~\cite{RefWorks:90}. Indeed, JavaScript from \url{optimizely.com} creates these cookies, although Optimizely states that they do not create third-party cookies~\cite{RefWorks:91}. \looseness=-1

	We also analyzed the first-party cookies on landing pages; see Fig.~\ref{fig:LP_hotspotfirstpartycookies}  in the appendix.  41 (\printpercent{42}{\crawlUniqueLocations}) hotspots create first-party cookies valid for various durations---e.g., over 5 years from  10 (\printpercent{10}{\crawlUniqueLocations}) hotspots, six months to five years from  42 (\printpercent{42}{\crawlUniqueLocations}) hotspots, and under six months from  41 (\printpercent{41}{\crawlUniqueLocations}) hotspots. Notable examples: Fossil has a 99-year valid cookie,  Fido has three cookies valid for 68--81 years,  CHU Sainte-Justine  has a 20-year valid cookie,  CIBC Bank  has a 19-year cookie, and Walmart has four cookies valid for 9--20 years. \looseness=-1
	
	\begin{figure}[ht]
	\vspace{-10pt}
\includegraphics[width=\textwidth, trim={2.5cm 13.7cm 1.9cm 5.6cm}, clip ]{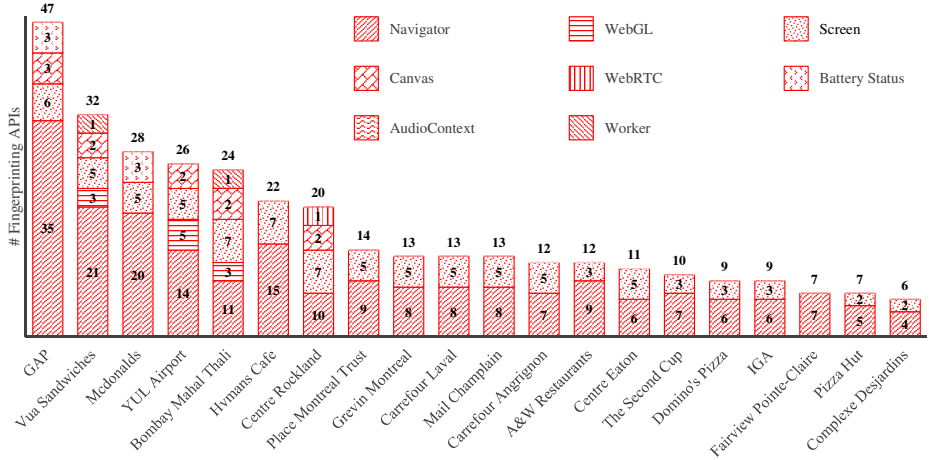}
		\vspace{-22pt}
		\caption{Unique number of fingerprinting APIs on captive portals (top 20). Note that for all fingerprinting statistics, we accumulate the distinct APIs as observed in all the datasets collected for a given hotspot.}
		\label{fig:CP_fingerprinting}
		\vspace{-18pt}
	\end{figure}

	\begin{figure}[ht]
		\vspace{-10pt}
		\includegraphics[width=\textwidth, trim={0.7cm 14.7cm 2.9cm 2.6cm}, clip ]{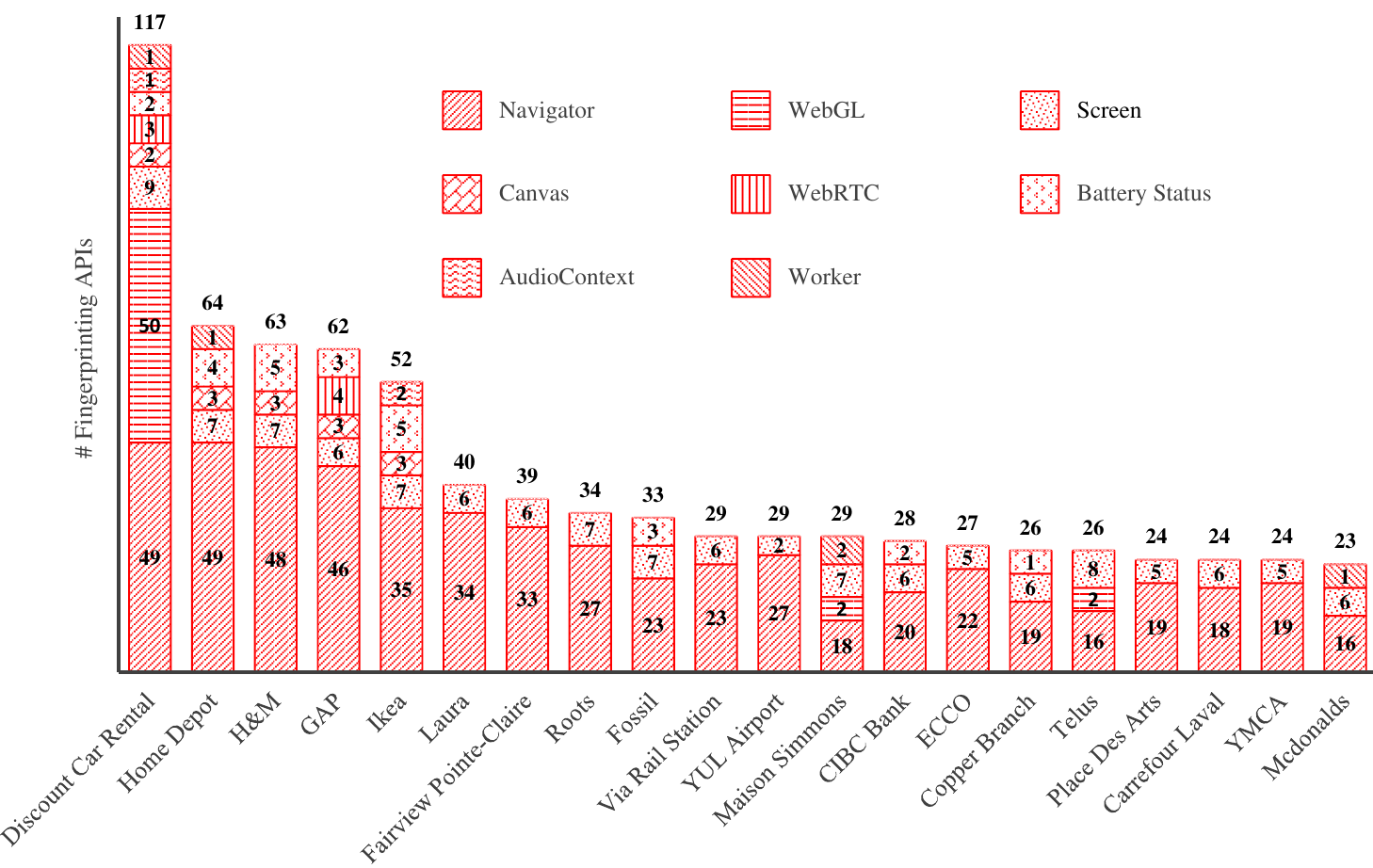}
		\vspace{-20pt}
	\caption{Unique number of fingerprinting APIs on landing pages (top 20)}
	\label{fig:LP_fingerprinting}
	\vspace{-18pt}
\end{figure}
	
\subsection{Device and Browser Fingerprinting}
		We analyzed fingerprinting attempts in captive portals and landing pages. We use Don't FingerPrint Me (DFPM~\cite{RefWorks:9}) for detecting known fingerprinting techniques, including the screen object, navigator object, WebRTC, Font, WebGL, Canvas, AudioContext, and Battery Status~\cite{RefWorks:7,nikiforakis2013cookieless,mowery2012pixel,olejnik2015leaking}. We use attribute and API interchangeably, when referring to fingerprinting JavaScript APIs. \looseness=-1
		
		\subhead{Captive portal}
	 24 (\printpercent{24}{\crawlUniqueLocations}) hotspots perform some form of fingerprinting. On average, each captive portal uses 5.9 attributes (max: 47 attributes, including 35 Navigator, 6 Screen, 3 Canvas, and 3 Battery Status); see Fig.~\ref{fig:CP_fingerprinting}. 
		We also found  10 (\printpercent{10}{\crawlUniqueLocations}) hotspots fingerprint user device/browser before explicit consent from the user, including GAP, McDonald's, and Place Montreal Trust, using 6--46 fingerprinting attributes. Our manual analysis of the GAP scripts reveals Font fingerprinting by checking the list of installed fonts using the side-channel inference technique described by Nikiforakis et al.~\cite{nikiforakis2013cookieless}. 
 Moreover, \readMacaddress{}  (\printpercent{\readMacaddress}{\crawlUniqueLocations}) captive portals fingerprint device MAC addresses.
 
	\subhead{Landing pages}
	 51 (\printpercent{51}{\crawlUniqueLocations}) hotspots perform fingerprinting on their landing pages. On average, each landing page fingerprints 19.4 attributes (max: 117 attributes, including 49 Navigator, 9 Screen, 2 Canvas, 3 WebRTC, 50 WebGL, 1 AudioContext, 1 Worker and 2 Battery Status); see Fig.~\ref{fig:LP_fingerprinting}. Prominent examples include the following. Discount Car Rental includes script from Sizmek Technologies Inc., which uses a total of 67 APIs (48 WebGL, 12 Navigator, five Screen, and two Canvas APIs). Manual analysis also reveals Font fingerprinting via side-channel inference~\cite{nikiforakis2013cookieless}; this script is also highly similar to FingerprintJS~\cite{fpjs}.
	Discount Car Rental also contains script from Integral Ad Science, which uses 41 attributes, including: 31 Navigator, seven Screen APIs, two WebRTC, and one AudioContext (cf.~\cite{RefWorks:7}). The navigator APIs are used to collect attributes such as the USB gamepad controllers using Navigator.getGamepads(), and list MIDI input and output devices using navigator.requestMIDIAccess. 
	H\&M and Home Depot host the same JavaScript that collects 42 attributes, including 34 Navigator, six Screen, and two Canvas APIs. 
	Laura has a script from PerimeterX that collects 27 attributes, including 21 Navigator and 6 Screen APIs; manual analysis of the source code reveals WebGL and Canvas fingerprinting. 

\section{CPInspector on Android} \label{sec:android}
In contrast to Windows, Android OS handles captive portals with a dedicated application. The Android Developers documentation and Android Source documentation omit details of how Android handles captive portals. Here we briefly document  the inner working of Android captive portals, and discuss our preliminary findings, specifically on tracking cookies on Android devices.

\begin{sloppypar}
\subhead{Android captive portal login app}
Using Android \texttt{ps} (Process Status), we observe that a new process named \texttt{com.android.captiveportallogin} appears whenever the captive portal is launched. The Manifest file for CaptivePortalLogin explicitly defines that its activity class will receive all captive portal broadcasts by any application installed on the OS and handle the captive portal. We observe that files in the data folder of this application are populated and altered during a captive portal session; we collect these files from our tests. 
\end{sloppypar}

\subhead{Capturing network traffic}
To capture traffic from Android apps, several readily-available VPN apps from Google Play can be used (e.g., Packet Capture, NetCapture, NetKeeper). However, Android does not use VPN for captive portals. 
On the other hand, using an MITM Proxy server such as mitmproxy (\url{https://mitmproxy.org/}) requires the server to run on a desktop environment, which would make the internet traffic come out of the desktop OS, i.e., the mobile device would not be visible to the hotspot. 
To overcome this, we set up a virtual Linux environment within the Android OS by using Linux Deploy (\url{https://github.com/meefik/linuxdeploy}), enabling us to run Linux desktop applications within Android with access to the core component of Android OS, e.g., Android OS processes, network interfaces, etc. We use Debian and mitmproxy on the virtual environment, and configure Android's network settings to proxy all the traffic going through the WiFi adapter to the mitmproxy server. The proxy provides us the shared session keys established with a destination server, enabling us to decrypt HTTPS traffic. We use tcpdump to capture the network traffic.

\subhead{Data collection and analysis}
We visited 22 hotspots and collected network traffic from their captive portals. First, we clear the data and cache of the CaptivePortalLogin app and collect data from a given hotspot. Next, we change the MAC address of our test devices (Google Pixel 3 with Android 9 and Nexus 4 with Android 5.1.1) and  collect data again without clearing the data and cache. 
From the proxy's request packets, we confirm that the browser agent correctly reflects our test devices, and the traffic is being originated from the CaptivePortalLogin app. Next, we analyze the data extracted from the app. The structure of the data directory is similar to Google Chrome on Android. We locate the  \verb!.\app_webview\Cookies! SQLite file in the data directory, storing the CaptivePortalLogin app's cookies.

\begin{figure}[ht]
	\vspace{-5pt}
		\includegraphics[width=\textwidth, trim={2.2cm 17.9cm 3cm 2.2cm}, clip ]{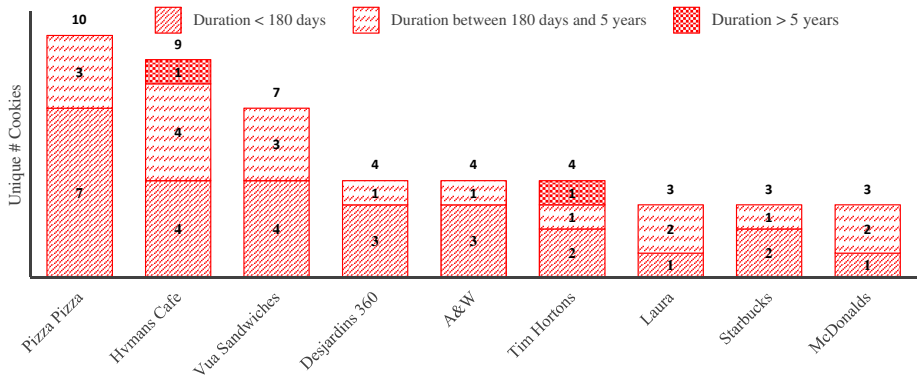}
        \centering
		\vspace{-35pt}
		\caption{Number of cookies stored on the Android captive portal app}
		\label{fig:androidcookies} 
		\vspace{-20pt}
	\end{figure}

We observe that 9 out of 22 hotspots store persistent cookies in the captive portal app; see Fig.~\ref{fig:androidcookies}. These cookies are not erased when the portal app is closed, or when the user leaves the hotspot. Instead, the cookies remain active as set in their validity periods, although they are unavailable to the regular browser apps. Prominent examples include: Tim Hortons inserts a 20-year valid cookie from \url{network-auth.com},
and Hvmans Cafe stores a 10-year valid cookie from Instagram.
In the captive portal traffic, we confirm that these cookies are indeed present and shared in subsequent visits, and follow the Same-Origin Policy. Hotspots can use these cookies to uniquely identify and authenticate user devices even when the device MAC address is dynamically changed; Tim Hortons hotspot uses its cookies for authentication. However, McDonald's did not authenticate the device even though the cookies were present but the MAC was new. \looseness=-1

\section{Privacy Policy and Anti-Tracking}	
\vspace{-8pt}
\subhead{Privacy policy}
We performed a preliminary manual analysis of privacy policy and TOS documents from hotspots that appear to be most risky. 
Roots states clearly in their privacy policy that they use SSL to protect PII, but their captive portal transmits a user's full name and email address via HTTP. 
Place Montreal Trust transmits the user's full name via HTTP, and they explicitly state that transmission of information over the public networks cannot be guaranteed to be 100\% secure. Nautilus Plus has a very basic TOS that omits important information such as the laws they comply with and privacy implications of using their hotspot. They state clearly that the assurance of confidentiality of the user's information is of great concern to Nautilus Plus, but they use HTTP for all communications, leaking personal information while they attempt to verify the customer's identity.  Their privacy policy is also inaccessible from the captive portal and omits any reference to WiFi. 

Dynamite and Garage are two brands of Groupe Dynamite. They transmit the user's email address via HTTP despite claiming to use SSL. Their privacy policy is inaccessible from the captive portal and omits any reference to the WiFi.
GAP explicitly mentions their collection of browser/device information, and they indeed collect 46 such attributes, \emph{before} the user accepts the hotspot's policies.\looseness=-1 

Although McDonald's tracks users in their captive portal (9 known trackers, 28 fingerprinting attributes), the captive portal itself lacks a privacy policy stating their use of web tracking.
Carrefour Laval and Fairview Pointe-Claire perform cross device tracking by participating in the Adobe Marketing Cloud Co-op~\cite{RefWorks:74}, where they may collect and share information about devices linked to the user. Two hotspots link the user’s MAC address to the collected personal information, including Roots, Bombay Mahal Thali.

Sharing the harvested data with subsidiaries and third-party affiliates is also the norm. Eight hotspots (including Hvmans Cafe, Fairview Pointe-Claire, and Carrefour Laval) state that PII may be stored outside Canada. Ten hotspots omit any information about the PII storage location, including Dominos's Pizza and Roots. However, five hotspots have their captive portal domain in the US, including Bombay Mahal Thali, Carrefour Angrignon, Domino's Pizza, Grevin Montreal, and Roots. 
 Three hotspots lack any privacy policy or a TOS document on their captive portals, including Laura, ECCO, and Maison Simmons.

	\subhead{Chrome vs.\ Firefox}
	 On captive portals, we found that the number of third-party tracking domains between Chrome and Firefox browsers differs by 5.3\% (Chrome: 353, including 79 known trackers; Firefox: 373, including 81 known trackers). 
	On landing pages, the difference is 7.7\% (Chrome: 2021, including 1317 known trackers; Firefox: 1865, including 1201 known trackers). Note that, landing pages generally host more dynamic content compared to captive portal pages; also, the Discount Car Rental hotspot lands on \url{msn.com} on Chrome and lands on the user's last visited page on Firefox. Moreover, due to various runtime errors, we could not test some hotspots in Firefox, including Carrefour Angrignon, Centre Rockland, Mail Champlain, and Centre Eaton. 
	
	\subhead{The same hotspot captive portal in different locations}
	\crawlMultipleLocations{} hotspots are measured at multiple physical locations. We stopped collecting datasets from different locations of the same chain-business as the collected datasets were largely the same. We provide an example where some minor differences occur: Starbucks' captive portal domain varies in the two evaluated locations (\url{am.datavalet.io} vs.\ \url{sbux-j2.datavalet.io}). However, the number of known trackers remained the same, while the number of third-parties increased by one domain. 
	Moreover, the \texttt{--sf-device} cookie validity increased from 17 days to 1 year, and the \texttt{--sf-landing} cookie was not created in the second location.

	\subhead{Effectiveness of privacy extensions and private browsing} To evaluate the effectiveness anti-tracking solutions
	against hotspot trackers, we collected traffic from both Chrome and Firefox in private browsing modes, and by enabling Adblock Plus~\cite{RefWorks:8} and Privacy Badger~\cite{RefWorks:10} extensions---leading to a total of six datasets for each hotspot. Then, we use the EasyList, EasyPrivacy, and Fanboy's lists to determine whether blacklisted requests or tracking cookies remain in the collected datasets; see
	Table~\ref{adblocingresults}. We only count the domain name of a tracker or advertiser when a request was sent, or a cookie was created. 
	
	\begin{table}[ht]
 \vspace{-15pt}
		\caption{The number of unique domains not blocked by our anti-tracking solutions.}
		\vspace{-5pt}
		\label{adblocingresults}
		\centering
	\begin{tabular}{l|c|c|c}
 & \textbf{\textbf{Ad Block Plus}} & \textbf{{Privacy Badger}} & \textbf{\textbf{Private Browsing}} \\ \hline
\textbf{Firefox} & 33 & 180 & 315 \\ 
\textbf{Chrome} & 117 & 212 & 356 \\ \hline
\end{tabular}%
	 \vspace{-20pt}
	\end{table}
 
\subhead{Hotspot trackers in the wild}
We measured the prevalence of trackers found in captive portals and landing pages, in popular websites---to understand the reach and consequences of hotspot trackers. We use OpenWPM~\cite{RefWorks:7} between Feb.\ 28--Mar.\ 15, 2019 to automatically browse the home pages of the top 143k Tranco domains~\cite{tranco-list} as of Feb.\ 27, 2019. We extracted the tracking persistent (validity $\geq$ 1 day; cf.~\cite{bujlow2017survey}) cookie domains from captive portals or landing pages. 
Then, we counted those tracking domains in the OpenWPM database; see Table~\ref{WhoTrackMe} in the appendix. For example, the \url{doubleclick.net} cookie as found in 4 captive portals and 30 landing pages, appears 160,508 times in the top 143k Tranco domains (mutiple times in some domains). 
Overall, hotspot users can be tracked across websites, even long time after the user has left a hotspot.

	\section{Conclusion}
	Many people across the world use public WiFi offered by an increasing number of businesses and public/government services. The use of VPNs, and the adoption of HTTPS in most websites and mobile apps largely secure users' personal/financial data from a malicious hotspot provider and other users of the same hotspot. However, device/user tracking as enabled by hotspots due to their access to MAC address and PII, remains as a significant privacy threat, which has not been explored thus far. Our analysis shows clear evidence of privacy risks and calls for more thorough scrutiny of these public hotspots by e.g., privacy advocates and government regulators. We will release our framework for easy replication and measurement in other parts of the world.
	
 	\section{Acknowledgement}
 	This work is supported by a grant from the Office of the Privacy Commissioner of Canada (OPC) Contributions Program.

	\bibliographystyle{splncs04}

\newpage	
 	\section* {Appendix} 
 		\begin{table}[ht]
	\vspace{-10pt}
		\caption{Sample of variations of the same third-party domain.} 
		\label{fig:Possible_Trackers}
		\resizebox{\textwidth}{!}{%
			\begin{tabular}{p{6.5in}c}
				\textbf{Third-Party Request-URL} & \textbf{Blacklisted} \\ \midrule
				https://www.google-analytics.com/r/collect?v=\&\_v=\&a=\&t=\&\_s=1\&dl=\&ul=\&de=\&dt= \&sd=\&sr=\&vp=\&je=\&\_u=\&jid=\&gjid=\&cid=\&tid=\&\_gid=\&\_r=1\&gtm=\&cd1= \&cd64= \&cd65= \&did= \&z= & Yes \\ \\
				https://www.google-analytics.com/j/collect?v=\&\_v=\&a=\&t=\&\_s=\&dl=\&ul=\&de=\&dt=\&sd=24-bit\&sr=\&vp=\&je=\&\_u= \&jid= \&gjid=\&cid=\&tid=\&\_gid=\&\_r=\&cd1=\&cd5=\&cd6=\&cd8=\&cd9=\&z= & No \\\bottomrule 
		\end{tabular}}
	\end{table}

\begin{table}[ht]
\label{WhoTrackMe}
\centering
\caption{Count of tracking domains from captive portals and landing pages in Alexa 143k home pages (top 10).}
\label{tab:my-table}
\begin{tabular}{|l|r|l|l|r|}
\cline{1-2} \cline{4-5}
\multicolumn{2}{|c|}{\textbf{Captive Portal}} & & \multicolumn{2}{c|}{\textbf{Landing Page}} \\ \cline{1-2} \cline{4-5} 
\multicolumn{1}{|c|}{\textbf{Tracker}} & \textbf{Count} & & \multicolumn{1}{c|}{\textbf{Tracker}} & \textbf{Count} \\ \cline{1-2} \cline{4-5} 
doubleclick.net & 160508 & & pubmatic.com & 326991 \\ 
linkedin.com & 48726 & & rubiconproject.com & 257643 \\ 
facebook.com & 37107 & & doubleclick.net & 160508 \\ 
twitter.com & 14874 & & casalemedia.com & 131626 \\ 
google.com & 13676 & & adsrvr.org & 116438 \\ 
atdmt.com & 5198 & & addthis.com & 83221 \\ 
instagram.com & 3466 & & demdex.net & 83160 \\ 
gap.com & 295 & & contextweb.com & 82965 \\ 
maxmind.com & 294 & & rlcdn.com & 75295 \\ 
gapcanada.ca & 64 & & livechatinc.com & 69919\\\cline{1-2} \cline{4-5} 
 \end{tabular}
\end{table}

\begin{figure}[!b]
\vspace{-10pt}        
\includegraphics[width=\textwidth, trim={1.9cm 16.5cm 2.5cm 1.9cm}, clip ]{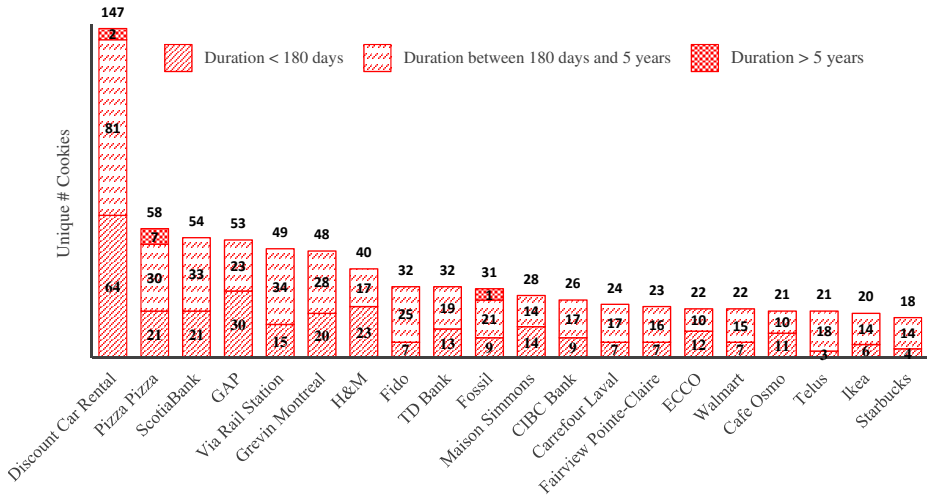}
		\vspace{-22pt}
    \centering
		\caption{Number of third-party cookies on landing pages (top 20)}
		\label{fig:LP_hotspotthirdpartycookies} 
		\vspace{-9pt}
	\end{figure}
	
	\begin{figure}[ht]
	\vspace{-10pt}
		\includegraphics[width=\textwidth, trim={2cm 16.1cm 3.2cm 2.6cm}, clip ]{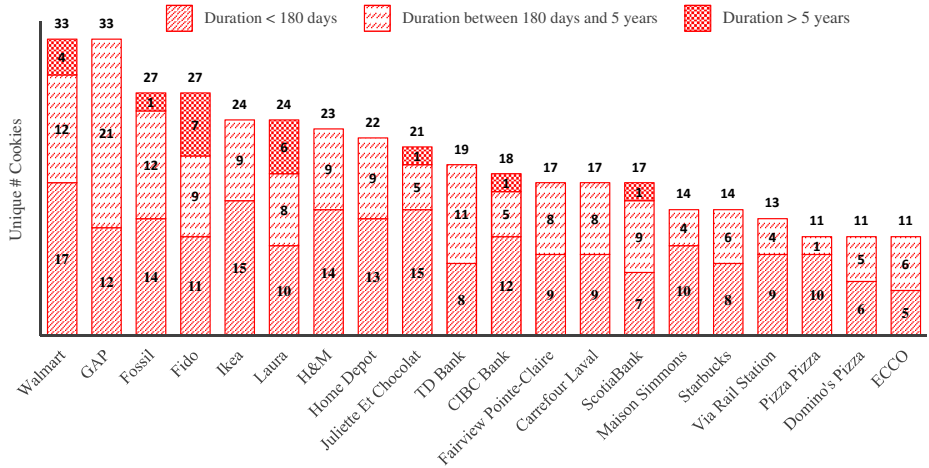}
        \centering

		\vspace{-22pt}
		\caption{Number of first-party cookies on landing pages (top 20)}
		\label{fig:LP_hotspotfirstpartycookies} 
		\vspace{-20pt}
	\end{figure}

\begin{table}[ht]
\caption{List of evaluated hotspots}
\begin{tabular}{@{}lc p{3in}@{}}
\toprule
\textbf{Category} & \textbf{Count} & \textbf{\ Hotspot Name} \\ \midrule
Cafe and Restaurant & 19 & \mbox{A\&W}, \mbox{Bombay Mahal Thali}, \mbox{Burger King}, \mbox{Cafe Osmo}, \mbox{Copper Branch}, \mbox{Domino's Pizza}, \mbox{Harvey's, Hvmans Cafe}, \mbox{Juliette Et Chocolat}, \mbox{Mcdonalds}, \mbox{Moose BAWR}, \mbox{Nespresso}, \mbox{Pizza Hut}, \mbox{Pizza Pizza}, \mbox{Starbucks}, \mbox{Sushi STE-Catherine}, \mbox{The Second Cup}, \mbox{Tim Hortons}, \mbox{Vua Sandwiches} \\
Retail business & 17 & \mbox{Canadian Tire},  \mbox{Dynamite}, \mbox{ECCO},  \mbox{Fossil}, \mbox{GAP}, \mbox{Garage}, \mbox{H\&M}, \mbox{Home Depot}, \mbox{IGA}, \mbox{Ikea}, \mbox{Laura}, \mbox{Maison Simmons}, \mbox{Michael Kors}, \mbox{Roots}, \mbox{SAQ}, \mbox{Sephora}, \mbox{Walmart} \\
Shopping Mall & 12 & \mbox{Atrium 1000}, \mbox{Carrefour Angrignon}, \mbox{Carrefour Laval}, \mbox{Carrefour iA}, \mbox{Centre Eaton}, \mbox{Centre Rockland}, \mbox{Complexe Desjardins}, \mbox{Fairview Pointe-Claire}, \mbox{Mail Champlain}, \mbox{Place Montreal Trust}, \mbox{Place Vertu}, \mbox{Place Ville Marie} \\
Bank & 5 & \mbox{CIBC Bank}, \mbox{Desjardins 360}, \mbox{RBC Bank}, \mbox{ScotiaBank}, \mbox{TD Bank} \\
Art and Entertainment & 4 & \mbox{Grevin Montreal}, \mbox{YMCA}, \mbox{Montreal Science Centre}, \mbox{Place Des Arts}  \\
Transportation & 3 & \mbox{Gare d'Autocars de Montreal}, \mbox{Via Rail Station}, \mbox{YUL Airport} \\
Telecom Kiosk & 2 & Fido, Telus \\
Car Rental & 1 & Discount Car Rental \\
Gymnasium & 1 & Nautilus Plus \\
Hospital & 1 & CHU Sainte-Justine \\
Hotel & 1 & Fairmont Hotel \\
Library & 1 & Westmount Public Library \\ \bottomrule
\end{tabular}
\label{hotspotscategory}
\end{table}

\end{document}